\documentclass[aps,prd,twocolumn,superscriptaddress,letterpaper,floatfix,nofootinbib,showpacs,preprintnumbers]{revtex4}
\usepackage{epsfig,units,color,ulem}

\newcommand{\at}{\alpha_{T}}
\newcommand{\zenith}{\theta_{zen}}
\newcommand{\zbar}{\bar{\theta}_{zen}}

\newcommand{\teff}{T_{\rm eff}}
\newcommand{\etal} {{\it et al.}}

\begin{document}
\preprint{FERMILAB-PUB-19-179-ND}

\title{Observation of seasonal variation of atmospheric
multiple-muon events in the NOvA Near Detector}

\begin{abstract}

Using two years of data from the NOvA Near Detector at Fermilab, 
we report a seasonal variation of cosmic ray induced
multiple-muon 
(N$_\mu$ $\geq$ 2) 
event rates which has an opposite phase 
to 
the seasonal
variation in the atmospheric temperature.  The strength of the seasonal multiple-muon variation is shown to
increase as a function of the muon multiplicity.  
However, no significant dependence of the strength of the seasonal
variation of the multiple-muon variation is seen as a function of the 
muon zenith angle, or
 the spatial or angular separation between the correlated muons.

\end{abstract}

\newcommand{\ANL}{Argonne National Laboratory, Argonne, Illinois 60439, 
USA}
\newcommand{\ICS}{Institute of Computer Science, The Czech 
Academy of Sciences, 
182 07 Prague, Czech Republic}
\newcommand{\IOP}{Institute of Physics, The Czech 
Academy of Sciences, 
182 21 Prague, Czech Republic}
\newcommand{\Atlantico}{Universidad del Atlantico,
Km. 7 antigua via a Puerto Colombia, Barranquilla, Colombia}
\newcommand{\BHU}{Department of Physics, Institute of Science, Banaras 
Hindu University, Varanasi, 221 005, India}
\newcommand{\UCLA}{Physics and Astronomy Department, UCLA, Box 951547, Los 
Angeles, California 90095-1547, USA}
\newcommand{\Caltech}{California Institute of 
Technology, Pasadena, California 91125, USA}
\newcommand{\Cochin}{Department of Physics, Cochin University
of Science and Technology, Kochi 682 022, India}
\newcommand{\Charles}
{Charles University, Faculty of Mathematics and Physics,
 Institute of Particle and Nuclear Physics, Prague, Czech Republic}
\newcommand{\Cincinnati}{Department of Physics, University of Cincinnati, 
Cincinnati, Ohio 45221, USA}
\newcommand{\CSU}{Department of Physics, Colorado 
State University, Fort Collins, CO 80523-1875, USA}
\newcommand{\CTU}{Czech Technical University in Prague,
Brehova 7, 115 19 Prague 1, Czech Republic}
\newcommand{\Dallas}{Physics Department, University of Texas at Dallas,
800 W. Campbell Rd. Richardson, Texas 75083-0688, USA}
\newcommand{\DallasU}{University of Dallas, 1845 E 
Northgate Drive, Irving, Texas 75062 USA}
\newcommand{\Delhi}{Department of Physics and Astrophysics, University of 
Delhi, Delhi 110007, India}
\newcommand{\JINR}{Joint Institute for Nuclear Research,  
Dubna, Moscow region 141980, Russia}
\newcommand{\FNAL}{Fermi National Accelerator Laboratory, Batavia, 
Illinois 60510, USA}
\newcommand{\UFG}{Instituto de F\'{i}sica, Universidade Federal de 
Goi\'{a}s, Goi\^{a}nia, Goi\'{a}s, 74690-900, Brazil}
\newcommand{\Guwahati}{Department of Physics, IIT Guwahati, Guwahati, 781 
039, India}
\newcommand{\Harvard}{Department of Physics, Harvard University, 
Cambridge, Massachusetts 02138, USA}
\newcommand{\Houston}{Department of Physics, 
University of Houston, Houston, Texas 77204, USA}
\newcommand{\IHyderabad}{Department of Physics, IIT Hyderabad, Hyderabad, 
502 205, India}
\newcommand{\Hyderabad}{School of Physics, University of Hyderabad, 
Hyderabad, 500 046, India}
\newcommand{\IIT}{Department of Physics,
Illinois Institute of Technology,
Chicago IL 60616, USA}
\newcommand{\Indiana}{Indiana University, Bloomington, Indiana 47405, 
USA}
\newcommand{\INR}{Inst. for Nuclear Research of Russia, Academy of 
Sciences 7a, 60th October Anniversary prospect, Moscow 117312, Russia}
\newcommand{\Iowa}{Department of Physics and Astronomy, Iowa State 
University, Ames, Iowa 50011, USA}
\newcommand{\Irvine}{Department of Physics and Astronomy, 
University of California at Irvine, Irvine, California 92697, USA}
\newcommand{\Jammu}{Department of Physics and Electronics, University of 
Jammu, Jammu Tawi, 180 006, Jammu and Kashmir, India}
\newcommand{\Lebedev}{Nuclear Physics and Astrophysics Division, Lebedev 
Physical 
Institute, Leninsky Prospect 53, 119991 Moscow, Russia}
\newcommand{\MSU}{Department of Physics and Astronomy, Michigan State 
University, East Lansing, Michigan 48824, USA}
\newcommand{\Duluth}{Department of Physics and Astronomy, 
University of Minnesota Duluth, Duluth, Minnesota 55812, USA}
\newcommand{\Minnesota}{School of Physics and Astronomy, University of 
Minnesota Twin Cities, Minneapolis, Minnesota 55455, USA}
\newcommand{\Oxford}{Subdepartment of Particle Physics, 
University of Oxford, Oxford OX1 3RH, United Kingdom}
\newcommand{\Panjab}{Department of Physics, Panjab University, 
Chandigarh, 160 014, India}
\newcommand{\Pitt}{Department of Physics, 
University of Pittsburgh, Pittsburgh, Pennsylvania 15260, USA}
\newcommand{\RAL}{Rutherford Appleton Laboratory, Science and 
Technology Facilities Council, Didcot, OX11 0QX, United Kingdom}
\newcommand{\SAlabama}{Department of Physics, University of 
South Alabama, Mobile, Alabama 36688, USA} 
\newcommand{\Carolina}{Department of Physics and Astronomy, University of 
South Carolina, Columbia, South Carolina 29208, USA}
\newcommand{\SDakota}{South Dakota School of Mines and Technology, Rapid 
City, South Dakota 57701, USA}
\newcommand{\SMU}{Department of Physics, Southern Methodist University, 
Dallas, Texas 75275, USA}
\newcommand{\Stanford}{Department of Physics, Stanford University, 
Stanford, California 94305, USA}
\newcommand{\Sussex}{Department of Physics and Astronomy, University of 
Sussex, Falmer, Brighton BN1 9QH, United Kingdom}
\newcommand{\Syracuse}{Department of Physics, Syracuse University,
Syracuse NY 13210, USA}
\newcommand{\Tennessee}{Department of Physics and Astronomy, 
University of Tennessee, Knoxville, Tennessee 37996, USA}
\newcommand{\Texas}{Department of Physics, University of Texas at Austin, 
Austin, Texas 78712, USA}
\newcommand{\Tufts}{Department of Physics and Astronomy, Tufts University, Medford, 
Massachusetts 02155, USA}
\newcommand{\UCL}{Physics and Astronomy Dept., University College London, 
Gower Street, London WC1E 6BT, United Kingdom}
\newcommand{\Virginia}{Department of Physics, University of Virginia, 
Charlottesville, Virginia 22904, USA}
\newcommand{\WSU}{Department of Mathematics, Statistics, and Physics,
 Wichita State University, 
Wichita, Kansas 67206, USA}
\newcommand{\WandM}{Department of Physics, College of William \& Mary, 
Williamsburg, Virginia 23187, USA}
\newcommand{\Winona}{Department of Physics, Winona State University, P.O. 
Box 5838, Winona, Minnesota 55987, USA}
\newcommand{\Wisconsin}{Department of Physics, University of 
Wisconsin-Madison, Madison, Wisconsin 53706, USA}
\newcommand{\Crookston}{Math, Science and Technology Department, University 
of Minnesota -- Crookston, Crookston, Minnesota 56716, USA}
\newcommand{\deceased}{Deceased.}
\affiliation{\ANL}
\affiliation{\Atlantico}
\affiliation{\BHU}
\affiliation{\Caltech}
\affiliation{\Charles}
\affiliation{\Cincinnati}
\affiliation{\Cochin}
\affiliation{\CSU}
\affiliation{\CTU}
\affiliation{\DallasU}
\affiliation{\Delhi}
\affiliation{\FNAL}
\affiliation{\UFG}
\affiliation{\Guwahati}
\affiliation{\Harvard}
\affiliation{\Houston}
\affiliation{\Hyderabad}
\affiliation{\IHyderabad}
\affiliation{\Indiana}
\affiliation{\ICS}
\affiliation{\IIT}
\affiliation{\INR}
\affiliation{\IOP}
\affiliation{\Iowa}
\affiliation{\Irvine}
\affiliation{\Jammu}
\affiliation{\JINR}
\affiliation{\Lebedev}
\affiliation{\MSU}
\affiliation{\Duluth}
\affiliation{\Minnesota}
\affiliation{\Panjab}
\affiliation{\Pitt}
\affiliation{\SAlabama}
\affiliation{\Carolina}
\affiliation{\SDakota}
\affiliation{\SMU}
\affiliation{\Stanford}
\affiliation{\Sussex}
\affiliation{\Syracuse}
\affiliation{\Tennessee}
\affiliation{\Texas}
\affiliation{\Tufts}
\affiliation{\UCL}
\affiliation{\Virginia}
\affiliation{\WSU}
\affiliation{\WandM}
\affiliation{\Wisconsin}

\author{M.~A.~Acero}
\affiliation{\Atlantico}

\author{P.~Adamson}
\affiliation{\FNAL}


\author{L.~Aliaga}
\affiliation{\FNAL}

\author{T.~Alion}
\affiliation{\Sussex}

\author{V.~Allakhverdian}
\affiliation{\JINR}

\author{S.~Altakarli}
\affiliation{\WSU}



\author{N.~Anfimov}
\affiliation{\JINR}


\author{A.~Antoshkin}
\affiliation{\JINR}

\affiliation{\Minnesota}



\author{A.~Aurisano}
\affiliation{\Cincinnati}


\author{A.~Back}
\affiliation{\Iowa}

% ...

\collaboration{The NOvA Collaboration}
\noaffiliation

\date{\today}
\maketitle

\section{Introduction}
\label{Introduction}
\par
This paper presents new measurements of the seasonality of underground 
multiple muons (N$_\mu$ $\geq$ 2) 
produced from cosmic ray showers in the atmosphere. 
Incoming cosmic ray nuclei interacting 
with the upper atmosphere produce a flux of 
pions ($\pi$), kaons (K), and other mesons at 
an altitude directly dependent upon the upper atmosphere density profile.  
These mesons either interact with the atmosphere to produce a hadronic 
cascade that contains additional mesons, or they decay to final states 
with 
muonic content.
The relative probability of each primary 
and secondary meson decaying, or having a 
strong interaction with the atmosphere,
depends on its energy and the density 
of the atmosphere near its production point.  
The density of the atmosphere depends upon 
many factors, with local temperature being the dominant one. 
The mean temperature of the upper atmosphere varies during the seasons, so 
the corresponding high energy cosmic muon rate is expected to vary.
The high energy muon flux increases
during the summer months due to the decrease in the density
of the upper atmosphere, which increases the 
probability that a meson will directly decay into a muon instead of
having a secondary strong interaction. 
Numerous underground 
detectors \cite{bib:barret,bib:sherman,bib:castagnoli2,
bib:fenton,bib:baksan,bib:macro,bib:amanda,bib:borexino,bib:lvd,
bib:ice,bib:ssw,bib:minosn,bib:minosf,bib:doublechooz,bib:daya}
at a variety of underground depths have measured this expected seasonal
variation via the flux of single-muon events. 
\par
The atmospheric particle showers produced by 
the interactions of cosmic ray nuclei 
produce muons of varying energies.  
The overburden associated with each underground 
detector will determine the minimum energy muon that can be 
observed. 
The highest energy muons usually come from $\pi$'s and K's produced in 
the first interaction of the primary cosmic ray in the atmosphere. 
The predominance of muons arising from daughters of the primary 
interaction is a consequence of the steeply falling power law for the 
cosmic ray energy spectrum
$\propto E^{-2.7}$, combined with Feynman scaling 
\cite{bib:feynman}
for the leading hadron in the primary interaction.
Observed underground single-muon events are produced by 
atmospheric showers in which the other muons, associated with the 
hadronic cascade, have either ranged out prior to reaching the 
detector or missed the detector due to the shower's angular 
divergence and extent at the detector location.
Thus it is expected that the 
observed muon in most single-muon events is 
the highest energy muon in the shower.
Multiple-muon events in an underground detector 
require one or more additional high energy muons at a small enough transverse
distance to be observed in the spatial limits of the detector. 
\par One important consideration in studying temperature effects in the 
atmosphere is the value of the critical energies for the $\pi$ and K.  
The critical energy is
defined as the energy for which the $\pi$~(K) interaction 
probability and decay probability
are equal.  Above the critical energy, more mesons interact before
they decay.  Below the critical energy, more mesons decay before they
interact.  The value for the $\pi$~(K) is 135$\,$GeV (850$\,$GeV).  Most
muons seen in shallow detectors (minimum energy at the 
Earth's surface $\left(E_\mu^{surface}\right)$ $<$ 100$\,$GeV) 
are 
from 
the decay of mesons below their
critical energies, which reduces the effect of temperature and density
fluctuations caused by seasonal effects, compared to higher energies
measured in deeper detectors.

 \par
The MINOS Near and Far Detectors 
observed a 
different seasonal 
variation for multiple-muon events than for single muons
\cite{bib:minos}.
The multiple-muon rate
was observed to unexpectedly increase during the 
winter months 
in the shallow underground Near Detector
($E_\mu^{surface} >$ 54$\,$GeV),
In the deeper Far Detector 
($E_\mu^{surface} >$ 730$\,$GeV),
the seasonal variation depended
upon the spatial separation of the muons in the event,
e.g. a winter maximum was seen for events with
muons within 4.5 m and
a summer maximum for events with muons separated more than 8 m.
\par
At low energy 
($E_\mu^{surface} \approx$ 1$\,$GeV),
muon decay 
plays a role
in seasonal effects.
We note that muon detectors located near the surface, such as the GRAPES
experiment ($E_\mu^{surface} >$ 1$\,$GeV),
measured a winter maximum for
their muon rate \cite{bib:grapes}.
The DECOR 
experiment 
($E_\mu^{surface} >$ 2$\,$GeV),
also measured a winter maximum for 
multiple muons on the surface \cite{bib:decor}.  
DECOR attributed their result to geometric effects
arising from altitude differences, but MINOS 
showed that 
at 
a depth of 225 meters water equivalent (mwe), 
the altitude differences were too small to explain the effect
\cite{bib:minos}.
\par
The goal of this analysis of NOvA data is to confirm and to further
investigate the seasonal 
effect that was measured in the MINOS experiment for multiple muons 
\cite{bib:minos},
with larger statistics, a simpler detector geometry, and looking at
the effect as a function of more observables.
This paper presents the multiple-muon rate observed in the NOvA Near 
Detector (ND) at Fermilab at a depth of 225$\,$mwe.  
The NOvA ND is at the same depth as the MINOS Near Detector but 
uses a different detector design. 
The muon rate in NOvA is measured using data from 
8 April 2015  to 16 April 2017,
representing two complete calendar years of exposure. This 
period does not coincide with the data presented by MINOS.
The strength of the multiple-muon seasonal rate variation is studied
using a Rayleigh power analysis, by looking for correlations with the
effective atmospheric temperature, and by fitting the rate to
a cosine function. The multiple-muon seasonal rate in NOvA is measured as 
a 
function of muon multiplicity and as 
a function of several geometric variables.

\section{The NOvA Near Detector and Muon and Temperature Data}
\label{sec:Detector}

The NOvA ND is located underground at a depth of 94$\,$m \cite{bib:nova}.
It was primarily designed to study neutrinos produced 
by
the Fermilab NuMI beam \cite{bib:numi}.
The detector is a segmented tracking calorimeter 
which
is constructed from planes of extruded polyvinyl chloride (PVC) cells
\cite{bib:talaga}.
Each NOvA  cell has a width of 3.8
$\,$cm, a 
depth of 5.9$\,$cm, and is 3.9$\,$m long.
The cells are filled with liquid scintillator \cite{bib:scint}
and the signal scintillation light 
is collected and transported to the readout by wavelength-shifting fiber 
which runs the length of each cell.
The light collected by the fibers is routed to avalanche photodiodes 
(APD)
and digitized.
Light producing a signal in the APD above a set threshold is recorded 
as a hit.  The detector and electronics are located in a climate
controlled environment which reduces one source of seasonal influence.

The detector consists of two parts: 
a fully active region and a muon ranger.  The 
active region contains 192 planes of cells.
Each plane is 3.9$\,$m by 3.9$\,$m in cross section.
The orientation of the planes alternates between vertical and 
horizontal views
around the beam to allow 
3D reconstruction.
The 
192 planes cover a longitudinal distance along the
NuMI beam of 12.75$\,$m. 
The muon ranger is located at the downstream end in the
beam direction.
It consists of 22 scintillator 
planes of size 3.9$\,$m horizontally by 2.7$\,$m
vertically. The muon ranger is  
2.85$\,$m long.  There are 10 steel planes of thickness 
10$\,$cm each
 interleaved with a pair of scintillator planes.
Together, the complete detector has 20,192 cells within the 214 planes.
The area at the top of the detector is $50\,\text{m}^2$ in the active 
detector
and $11\,\text{m}^2$ in the muon ranger.

Cosmic rays are recorded in the NOvA ND with an “activity” trigger which 
requires at least 10 hits on at least 8 planes in total with at least 3 
planes hit in each of the two views. In addition, there must be at least 5 
planes with hits in a window of 6 sequential planes.  The typical activity
trigger rate is 39 triggers/s. 
Each trigger causes a readout of 50 or $100\,\mu$s of data which fully 
encompasses the hits which satisfy the triggering condition. This hit data 
has a single hit timing resolution of 5-10$\,$ns. In this analysis, tracks 
registering in the detector with temporal 
separation of less than 100$\,$ns are considered to be correlated and part 
of 
a multiple-muon event.
Data overlapping with the NuMI beam spill was not used for this analysis. 
Cosmic muon reconstruction is performed 
using a Hough Transform \cite{bib:hough} which finds hits that line up in
each view.  The two views are then matched
to produce a 3D reconstructed track.   

In order to reduce the number of misreconstructed events to a negligible
level, additional analysis selection criteria were applied to the
events.  
NOvA monitors the quality of its data continuously and only those data
meeting publication quality standards were used in this analysis.
Reconstructed track directions along the planes of the detector were 
discarded
because many resulted from bad matching of 2D tracks.
This was done by selecting  the direction cosines in the X and Z
directions; $| \cos\theta_z | \ge $ 0.02 and 
 $\cos\theta_x\ge 0.02$ or $\cos\theta_x\le 0$. 
 In addition, to remove short tracks consistent with electrons
from bremsstrahlung above the detector, we impose a throughgoing
requirement by demanding the first and last hit on all tracks be within
50$\,$cm of the detector edges.  This selection
removes stopping muons 
which are
the 2\% of incoming muons with the lowest energies.

Using a Monte Carlo simulation  
(MC) based on the CRY simulation \cite{bib:cry}, 
the reconstruction efficiency after all selection criteria was 69\%.
This was consistent with the result that 73\% of all activity triggers
gave at least one selected muon.
The inefficiency comes
from both the 10-hit requirement and reconstruction difficulties for
steep tracks.  The efficiency estimate is not important for the 
rest of the analysis since it does not depend on time during the year.
A two-muon
 simulation was developed using the single-muon 
simulation and randomly 
placing a second parallel muon in the detector. 
Both muons were reconstructed and passed the analysis
criteria with an efficiency of 37\%.
The two-muon efficiency was reduced some due to confusion
when 2D tracks overlapped in one view.
A visual inspection of several thousand triggers showed the 
impurity from triggers not containing muons  
(before reconstruction) to be below 1\%.
There was agreement 
of the distributions of track positions and angles
between the data and 
simulation 
\cite{bib:stefthesis}.  
Other than as a 
check on the validity of the reconstruction, a simulation was not 
used in the analysis presented here.
\par
\begin{figure}[thbp]
  \begin{center}
    \includegraphics[width=0.48\textwidth]{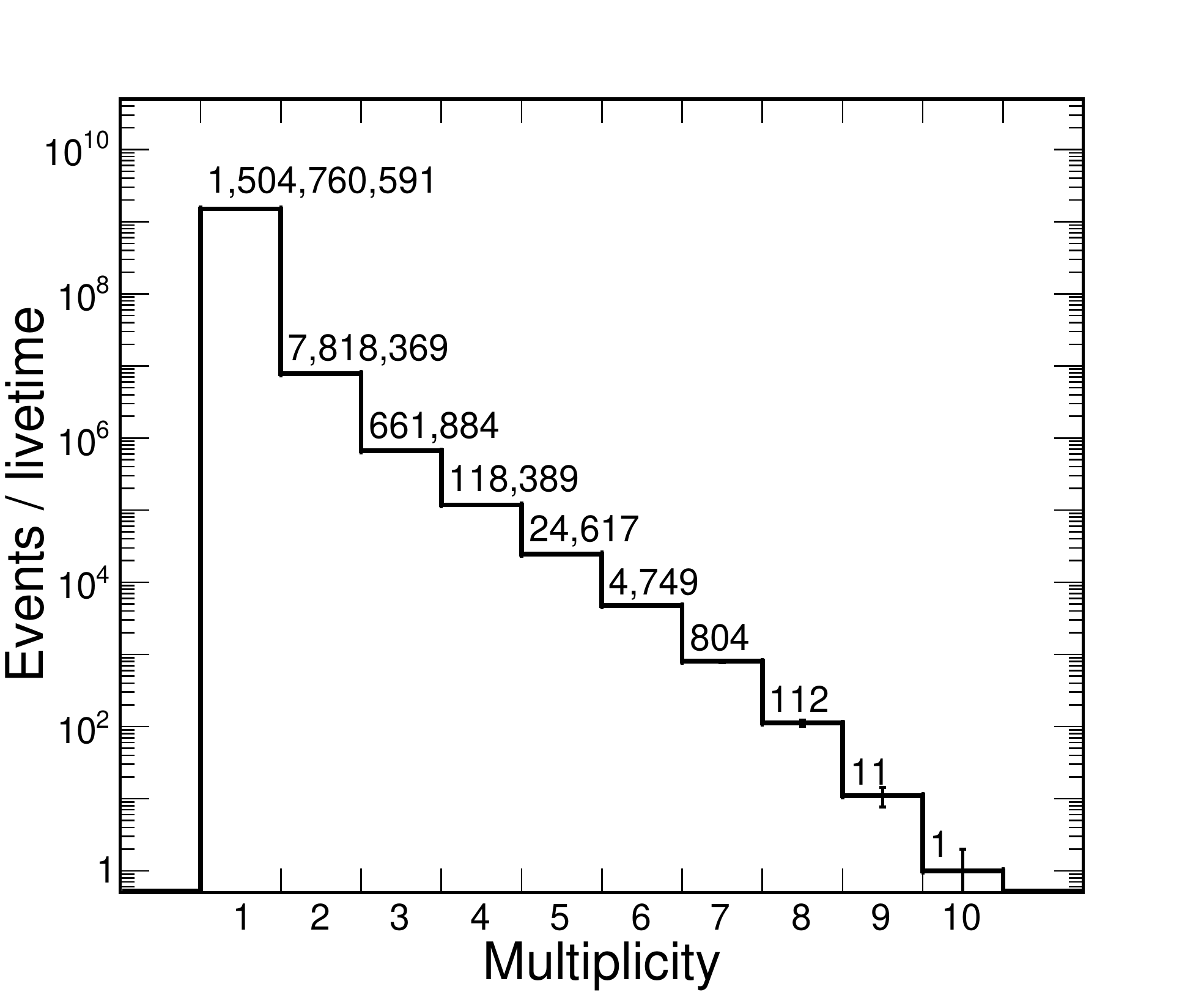}
  \end{center}
  \caption{Observed multiplicity distribution for single- and
multiple-muon events
in the NOvA ND.  The livetime for this exposure was
55.29$~\times~{\rm{10}^6}\,$s. 
Note that the
vertical axis in the figure is shown on a logarithmic scale.  
  \label{fig:multiplicity}}
\end{figure}

 The reconstructed track multiplicity for multiple-muon events in
NOvA is shown in 
Fig.~\ref{fig:multiplicity}.  
The maximum reconstructed
multiplicity event found in our sample is 10 muons.  
In this paper, the multiplicity always refers to the observed
multiplicity.  We do not correct 
for
muons within air showers that reach the depth of the detector but miss it 
laterally.  Thus the muon multiplicity is a detector 
(acceptance) dependent quantity.
\par
\begin{figure}[thbp]
  \begin{center}
    \includegraphics[width=0.48\textwidth]{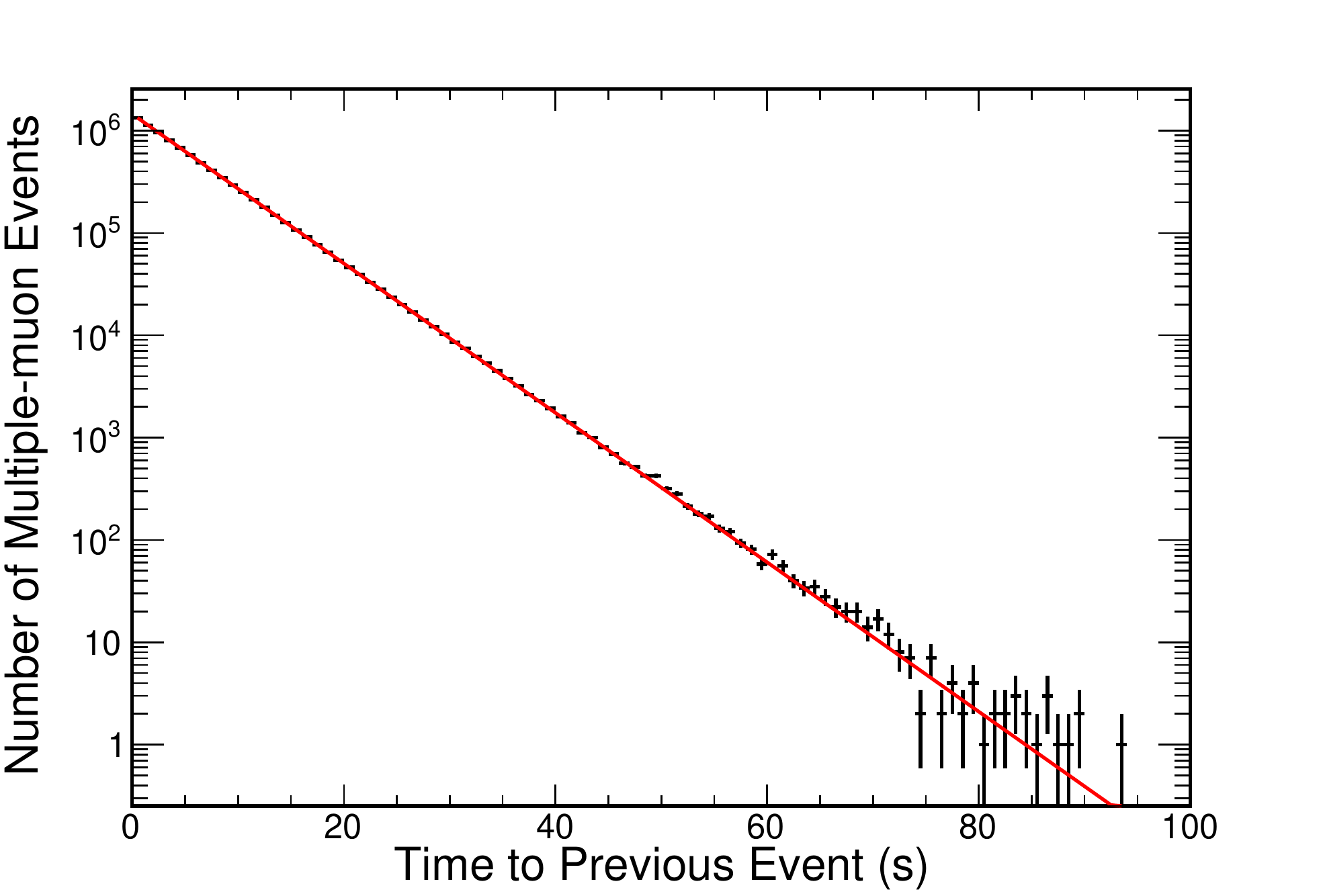}
  \end{center}
  \caption{Time between multiple-muon events in the NOvA
ND.  An exponential fit is shown, as expected for random
events with no correlations.
 The mean rate from this fit is 0.17 s$^{-1}$.
 \label{fig:time}}
\end{figure}

 The total elapsed time for this period is
63.85$~\times~{\rm{10}^6}\,$s.  Event rates were calculated during
periods in which data was recorded that were up to an
hour long.  Rates during longer periods were calculated using
the number of observed events and the corresponding livetime.
The total detector livetime was 55.29$~\times~{\rm{10}^6}\,$s
representing a livetime fraction of 86\%.   The livetime was not
uniformly distributed, but there was ample statistics to calculate a rate 
during
every month.
The time between multiple-muon events during periods of
livetime is shown in
Fig.~\ref{fig:time}.  The distribution
drops according to a power law
over several orders of magnitude, as
expected for
random uncorrelated events. 
\par Atmospheric temperature
data is provided four times per day
by the European Center for Medium-Range Weather
Forecast (ECMWF) at 37 pressure levels, ranging from 1$\,$hPa to 
1,000$\,$hPa, corresponding to altitudes up to 50$\,$km  
\cite{bib:ecmwf}.  ECMWF provides
interpolated temperature values on the corners of a 
grid, whose latitude and longitude values range from (41.25$^{\circ}$
N, 87.75$^{\circ}$ W) to (42.00$^{\circ}$ N, 88.50$^{\circ}$ W) with a
0.75$^{\circ}$ increment in each direction.  This area well matches
the production site for most of the muons reaching the NOvA
ND at 41.50$^{\circ}$ N, 88.16$^{\circ}$ W
\cite{bib:stefthesis,bib:norman}.  These temperature values are
used to construct $\teff$, which is their average weighted over
the altitude for single-muon production \cite{bib:grashorn}.

\section{Seasonal Analysis}
\par
\begin{figure}[thbp]
  \begin{center}
    \includegraphics[width=0.48\textwidth]{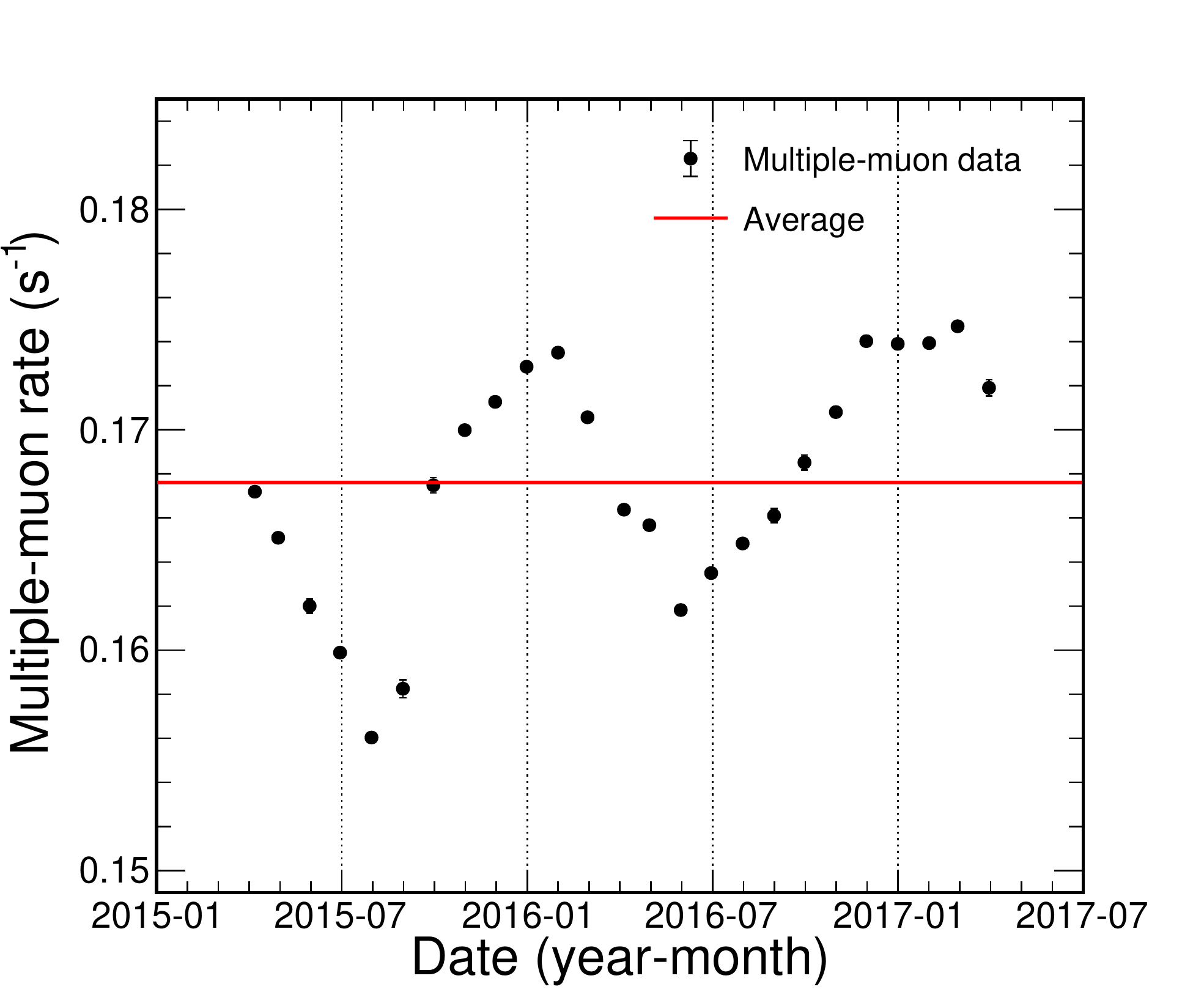}
  \end{center}
  \caption{Rate of multiple muons in the NOvA ND
as a function of month and year.
  \label{fig:monthly}}
\end{figure}

\begin{figure}[thbp]
  \begin{center}
    \includegraphics[width=0.48\textwidth]{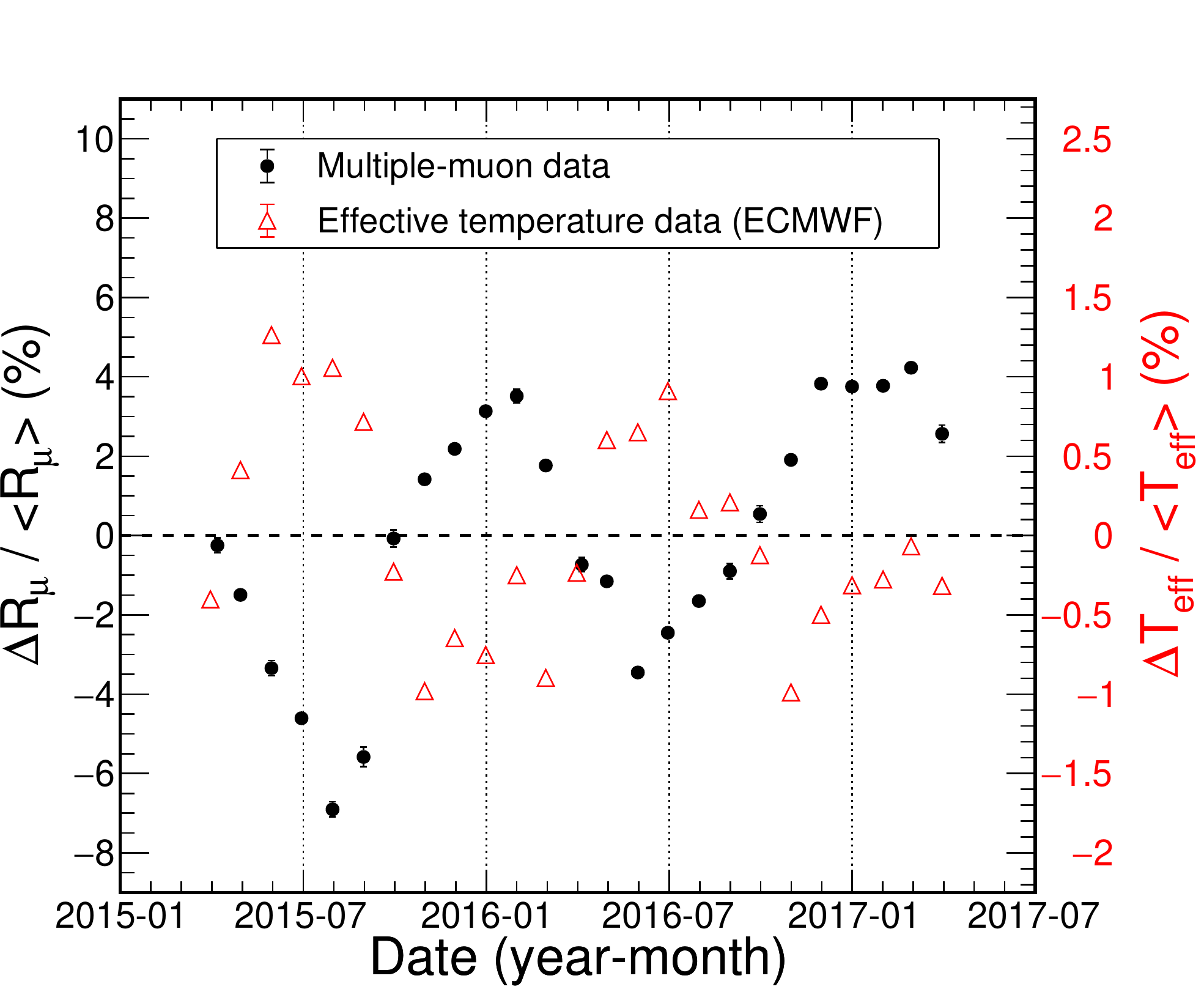}
  \end{center}
  \caption{
 Rate variation of
the multiple muons as a function of month shown
with the variation in the effective atmospheric temperature for single 
muons above the NOvA ND.  The mean values are $\langle R_\mu \rangle$ = 
0.168 $s^{-1}$
and $\langle \teff \rangle$ = 222 K.
  \label{fig:teff}}
\end{figure}

 
The observed rate of multiple muons $\left(R_\mu\right)$ is shown using 
bins 
corresponding to one month in time in
Fig.~{\ref{fig:monthly}}.  
A clear seasonal variation is observed.
 The size of the winter/summer rate change
differs between the two years of data.  A number of consistency 
checks showed that there was no difference in detector performance
affecting this analysis
during those two years \cite{bib:stefthesis}.  
The effective
temperature calculated at the production altitude for single muons 
above the NOvA ND was calculated in a similar way as in 
reference~\cite{bib:minos}.
The monthly values of $\Delta(\teff) /\langle \teff \rangle$ 
and $\Delta(R_\mu) / \langle R_\mu \rangle$
are
shown in Fig.~{\ref{fig:teff}}.  An anticorrelation between these
two quantities is evident. 
\par 
Since the frequency we were testing is well known, a 
frequency
analysis using the  Lomb-Scargle method
\cite{bib:lomb} was performed on the multiple-muon data as
a consistency check.  The 
highest power was found at a frequency corresponding to a 
year \cite{bib:stefthesis}.  A strong seasonal effect is
apparent in Fig.~{\ref{fig:monthly}}.
To further study this variation as a function of several observables,
it was necessary to select an
{\it a priori} 
 way to quantify 
the sign and strength of the effect.
We chose three complementary methods:
1) a Rayleigh power analysis, 
2) the correlation coefficient $\at$ of the rate
with effective temperature,
and 
3) comparison of the rate change to a cosine function.
MINOS has shown a seasonal multiple-muon effect with an opposite
phase to that for the single muons \cite{bib:minos}, however we
extended the previous qualitative 
 analysis with these methods.  
Each method has some advantages and disadvantages
in this context.

\subsection{Rayleigh analysis}
\par The Rayleigh analysis uses the binned 
Rayleigh power ($P_R$), which is 
defined as:
\begin{equation}
P_R  =
\frac{ 
\left\{\sum^n_{i=1} x_i \sin(\omega t_i)\right\}^2 +
\left\{\sum^n_{i=1} x_i \cos(\omega t_i)\right\}^2 
}{N},
\end{equation}
where $N$ is the total number of events, $n$ is the number of bins,
$x_i \equiv x(t_i)$ is the number of events in each bin, $\omega~=~2 
\pi/$(1~year) 
is 
the angular
frequency,
and $t_i$ is the time of the center of each bin.
The 
Rayleigh power can be thought of as the deviation from
the origin for a random walk of $N$ steps.  Since the frequency
is known, it gives an absolute probability that unseasonal 
data would give the observed power.  This method is compromised by 
gaps
in the data for small bin sizes, but for monthly or even weekly bins 
 there are no gaps.  However, to compare the size of the
power for subsamples of the data, the number of events in each
subsample needs to be identical.
It is not useful, for example, for comparing the power of 
different multiplicities because the sample sizes widely differ.  The 
binning in time is chosen to have a negligible 
effect on
the calculation of $P_R$.
The chance probability that the obtained value of the Rayleigh power does 
not
come from a random flat distribution is $1-e^{-P_R}$.  All probabilities
obtained in this analysis are near unity, but the value of $P_R$ itself is
used to see if there are trends as a function of interesting variables.

\par A calculation of the Rayleigh power using the data in 
Fig.~\ref{fig:monthly} gives a value $P_R$ = 3665.
The probability that this is the result of non-seasonal random data
is $e^{-3665}$, which is negligible.

\subsection{Correlation coefficient}
\par
Seasonal variations for single muons have been studied with a
correlation coefficient $\at$ defined by \cite{bib:grashorn}
\begin{equation}
\frac{\Delta R_{\mu}}
{\ \langle R_{\mu} \rangle}=\at\frac{\Delta \teff}
{\langle \teff \rangle},
\label{eq:alpha}
\end{equation}
where $\langle R_{\mu}\rangle$ is the mean muon event rate
for the complete observation period, and corresponds to 
the rate for an effective atmospheric temperature equal to 
$\langle \teff \rangle$.
The magnitude of the temperature coefficient $\at$
 is dependent on the muon energy at production and hence the 
depth of the detector. The effective temperature
$\teff$ is a weighted average of temperature measurements over 
the region of the atmosphere where muons originate \cite{bib:grashorn}. 
The value of $\teff$ tracks the actual temperatures at 37 altitudes 
calculated on a 6-hour basis.  This temperature is correlated with the density 
of the atmosphere and hence the competition between interaction and decay for 
$\pi$'s and K's as they traverse the varying density atmosphere.  As a 
consequence of the steeply falling energy distribution of cosmic ray 
primaries, only considering hadrons in the
first interaction is a good 
approximation for single muons.  A theoretical formula for $\at$
for single muons derived in reference \cite{bib:grashorn} gives a value 
that is always positive.  

For this multiple-muon analysis, a limitation is that 
$\teff$ in Eq.~\ref{eq:alpha}
 has been calculated by weighting the vertical temperature distribution with the 
interaction length of the primary cosmic ray together with the lifetime of a 
secondary hadron produced in the first interaction [30].  However,  
the 
seasonal behavior of the rate for multiple-muon events is not expected to be 
precisely represented by a simple formula due to the many competing 
effects such as non-leading mesons from the first
interaction, and mesons
 from secondary interactions, etc.    Multiple muons observed underground may 
predominantly result from hadrons produced in secondary interactions 
or those further into a hadronic shower. The calculation 
of $\teff$ used above is a poorer approximation in the determination of $\at$
than for single muons. However, the gradient of temperature variations in the 
atmosphere is 
fairly smooth in both winter and summer, so $\teff$ may be useful in tracking 
the multiple-muon effective temperature variation as a function of 
date and is used in the analysis below.
Using the data in Fig.~\ref{fig:teff}, we find
$\at$ = -4.14 $\pm$ 0.07.  
The quoted uncertainty comes from the fit and does not include
the systematic uncertainties discussed below.

\subsection{Cosine fit}
\par Our third measure of the strength of the 
seasonal variation is the
amplitude of a fit of the data to a cosine function.  The fitting 
function used is 
\begin{equation}
f(t) = V_0 + V \cos(\omega t + \phi),
\end{equation}
and the amplitudes $V$ are compared in the next section.
While
temperatures are predictably warmer in the summer and colder in the
winter, the variation does not typically follow a cosine function,
so any fit will necessarily be poorly described by that function.
In fact the difference between the two years in Fig.~\ref{fig:monthly} is 
larger 
than the differences seen in reference~\cite{bib:minosn}.
Nevertheless, we find such a fit to be a 
useful way to parametrize
some of the data.  
The fits were performed on the data binned according to the month of the 
year in which the data were recorded.

\par
\begin{figure}[thbp]
  \begin{center}
    \includegraphics[width=0.48\textwidth]{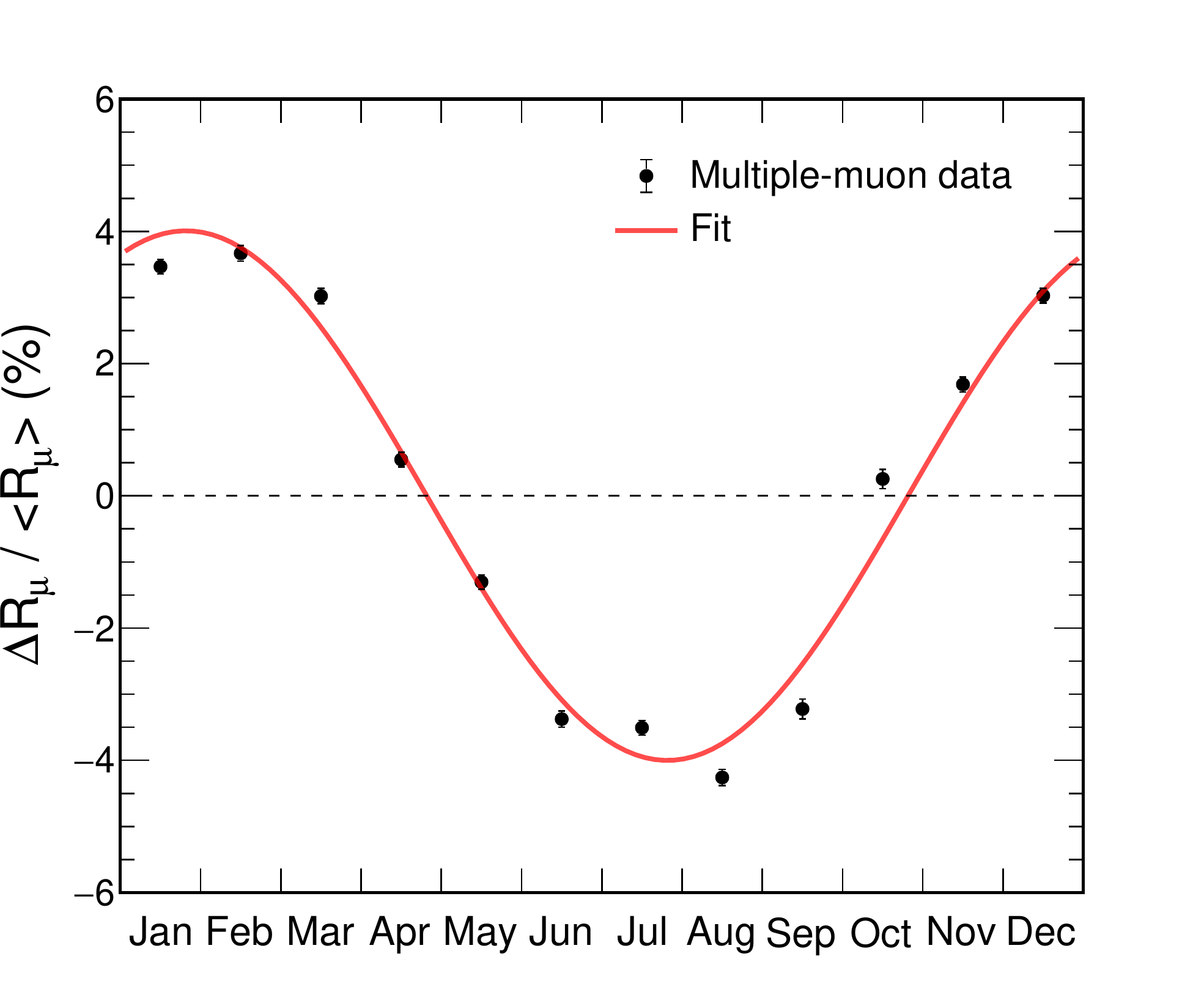}
  \end{center}
  \caption{Percentage rate variation of multiple muons in the NOvA ND
as a function of month of year.
  \label{fig:monthofyear}}
\end{figure}

 Averaging over the two years of NOvA data, we show the multiple-muon
rate as a function of the
 month of year in Fig.~\ref{fig:monthofyear}.  
That
distribution is more sinusoidal than the rate as a function of time,
as had been observed previously \cite{bib:minos}.  We perform the fit to
the data in  Fig.~\ref{fig:monthofyear} and obtain 
$V_0$ = 0.0 $\pm$ 0.1 \%,
$V$ = 4.1 $\pm$ 0.2 \%, and $\phi$ = -0.43 $\pm$ 0.05 radians.  
The  uncertainty is only that from the fit.
This 
value 
of
the phase corresponds to a maximum multiple-muon rate near 25 January
and a minimum near 26 July.  In all subsequent fits
we set $\phi$ = -0.43 radians.
The value of $V_0$ in every fit is consistent with zero.

\section{Studies of  multiple-muon observables in the NOvA ND}
\label{sec:4}
\par The minimum muon energy needed to reach the NOvA ND
through
the overburden depends on the zenith angle ($\theta_{zen}$)
and is 
approximately 
proportional to $\sec{\theta_{zen}}$.  The highest
energy muons come from the highest energy primary cosmic rays.  Since
the cosmic ray energy spectrum is a steeply falling function,
a test of the seasonal variation as a function of zenith 
angle $\zenith$ can be used to look for an energy dependence.  
\par
\begin{figure}[thbp]
  \begin{center}
    \includegraphics[width=0.48\textwidth]{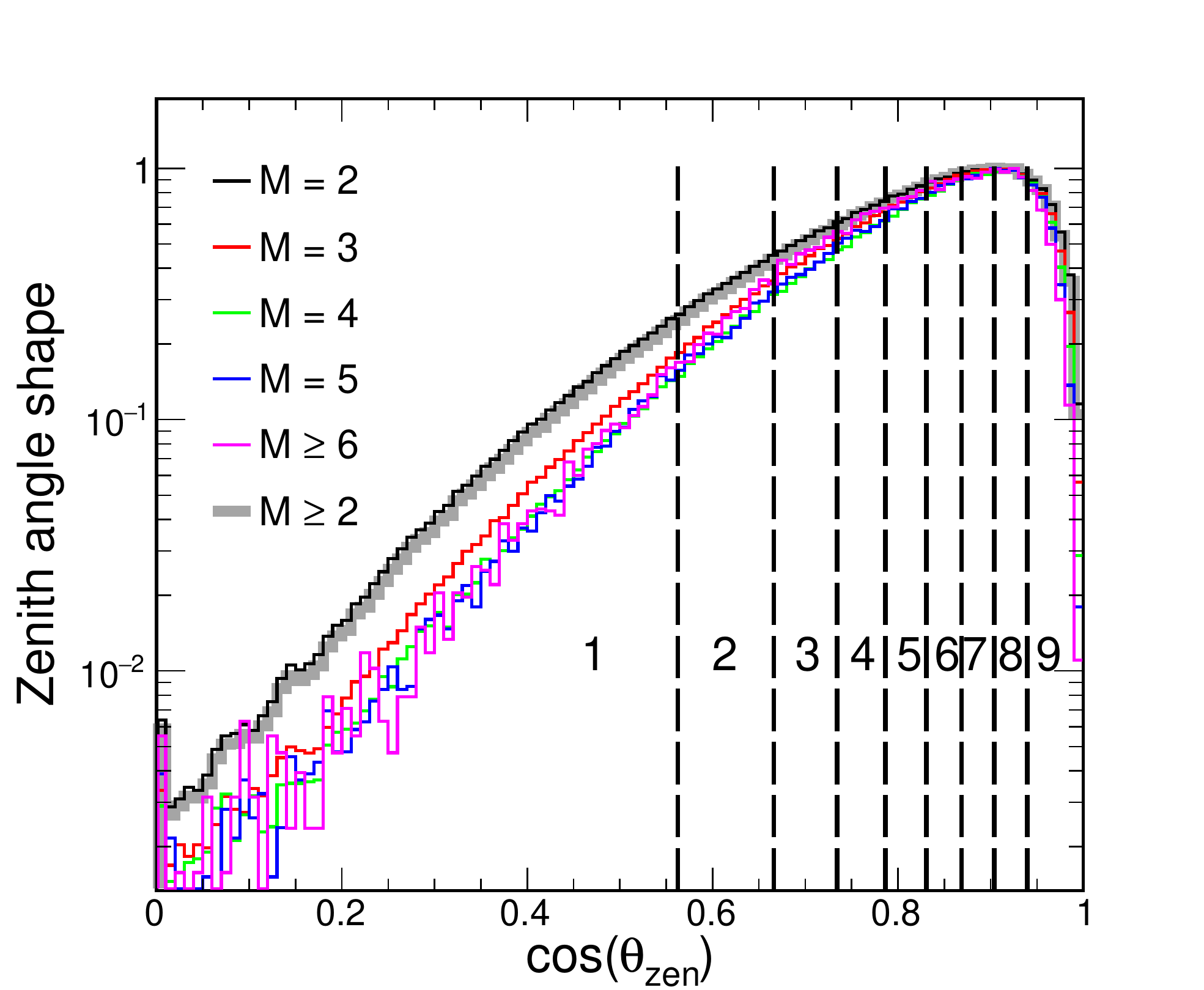}
  \end{center}
  \caption{Zenith angle distribution for each track in a
multiple-muon event in the NOvA ND. 
Multiplicity distributions are normalized to have the maximum equal to 1 
(one).  The regions marked 1-9 have equal statistics.
\label{fig:zenith}}
\end{figure}

 The zenith
angle distribution for each track in a multiple-muon event
is shown in Fig.~\ref{fig:zenith}.  The distribution is divided into
nine equal data sets which were used to calculate the Rayleigh Power,
$\at$, and the amplitude $V$ of the cosine fit.  Those values for the 
nine
regions are shown in Table~\ref{tab:zenith}.  
There do not appear
to be any differences between the seasonal variation 
of multiple-muons at low
and high zenith angles.

\begin{table}[tbh]
\begin{tabular}{cccccc}
\hline
Sample & $\cos \theta_{zen}$ & $P_R$ & $\at$ & $V$ (\%)& 
Tracks 
\\ 
& & & $\pm$ 0.1 & $\pm$ 0.1 & \\
\hline \hline
1 & $<$ 0.562& 1475 & -3.6  & 3.8  & 1,960,354\\ \hline
2 & 0.562-0.666& 1624 & -3.8  & 4.0  & 1,960,477\\ 
\hline
3 & 0.666-0.734& 1732 & -3.8  & 4.0  & 1,960,777\\ 
\hline
4 & 0.734-0.787 & 1778 & -3.7 & 4.0 & 1,960,676 \\ 
\hline
5 & 0.787-0.830 & 1715 & -3.5  & 3.8 & 1,960,327 \\ 
\hline
6 & 0.830-0.869 & 1807 & -3.9 & 4.0  & 1,960,581 \\ 
\hline
7 & 0.869-0.904 & 1667 & -3.5  & 3.7  & 1,960,739 \\ 
\hline
8 & 0.904-0.939 & 1562 & -3.8  & 3.8  & 1,961,248 \\ 
\hline
9 & $>$ 0.939 & 1563 & -4.2  & 4.2 & 1,957,395 
\\ 
\hline
\end{tabular}
\caption{\label{tab:zenith} 
Zenith angles are calculated for each 
track in a multiple-muon event.  
Measurements of the seasonal
variation are shown for
nine regions of  $\cos \theta_{zen}$.  
The uncertainties on $\at$ and $V$ are from the fit.
}
\end{table}

\par In the MINOS Far Detector, a difference in the seasonal variation
of multiple-muon events was seen as a function of separation
distance between the muons \cite{bib:minos}.  In the smaller 
MINOS Near Detector
the same variation was not seen.  Since the typical transverse
momentum $\left(p_t\right)$ for
a hadron in a hadronic interaction is 300$\,$MeV/c, the distance
between muons in the detector may decrease with increasing
primary and muon energies.  Multiple scattering in the overburden
also affects this distance, but multiple scattering is smaller
for larger muon energies.  The track separation in NOvA is
calculated by taking the perpendicular distance between every
pair of tracks in a multiple-muon event
\begin{equation}
\Delta L = \cos{\zbar} \times \sqrt{(\Delta X)^2 + (\Delta Z)^2},
\end{equation}
where $X$ and $Z$ are the horizontal detector coordinates of each
track at the top of the detector 
and $\zbar$ is the average zenith angle of the
two tracks.  
\par
\begin{figure}[thbp]
  \begin{center}
    \includegraphics[width=0.48\textwidth]{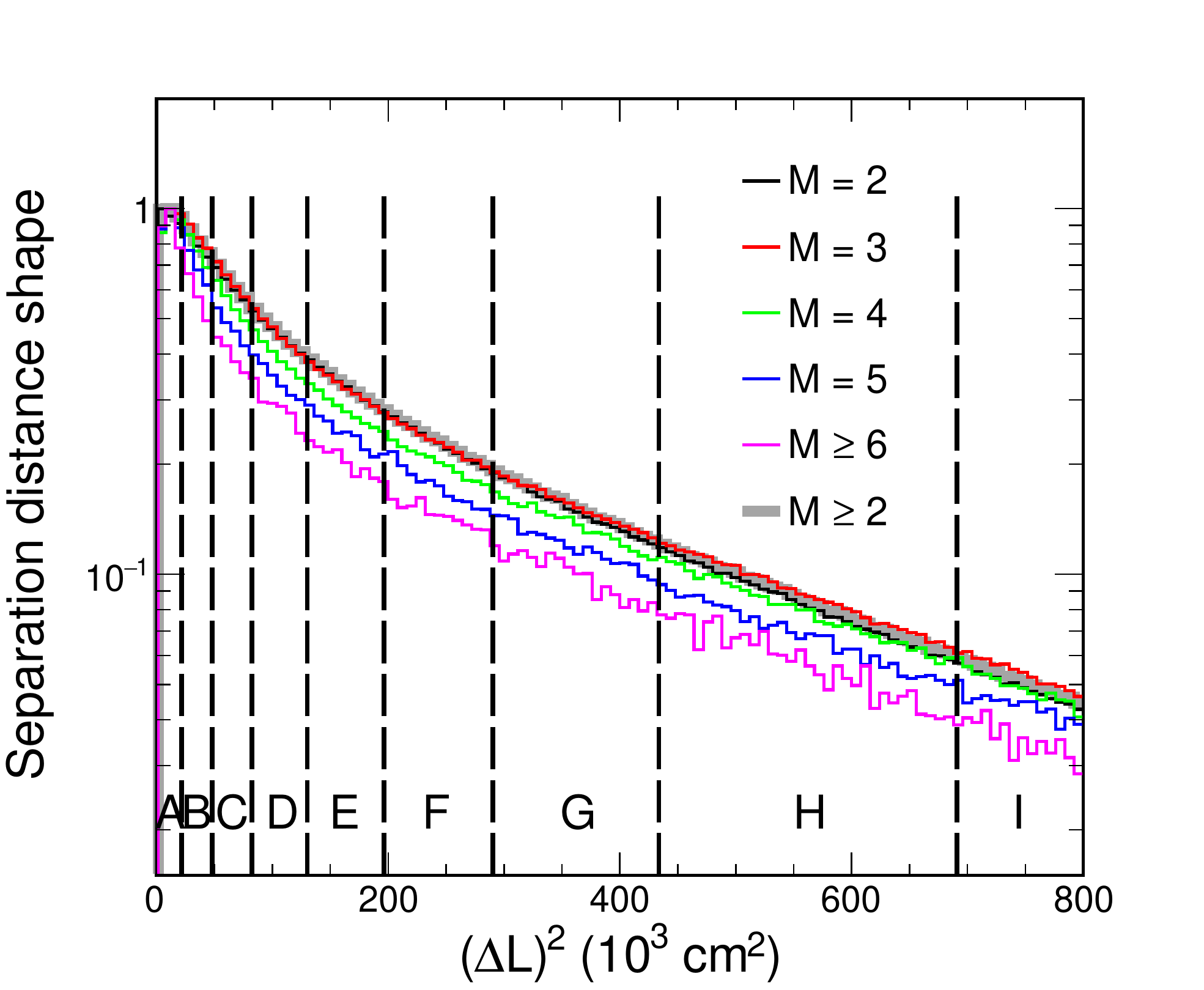}
  \end{center}
  \caption{Square of the separation $\Delta L$ between each track in each
muon pair in 
a multiple-muon
event in the NOvA ND.  
Multiplicity distributions are normalized to have the maximum equal to 1 
(one).  The regions marked A to I have equal statistics.
  \label{fig:separation}}
\end{figure}

 The square of the track separation $\Delta L$ is shown in
Fig.~\ref{fig:separation}.   Nine equal-statistics regions (A...I) of 
track separation are defined with limits found in 
Table~{\ref{tab:separation}}.
While the first and last bins show larger values of $P_R$, $\at$ and $V$, 
there does not appear
to be any trend showing a difference between the seasonal variation 
of multiple-muons at large and small separation.

\begin{table}[thb]
\begin{tabular}{crrrcr}
\hline
Sample & $(\Delta L)^2$ (10$^3$ cm$^2$) & $P_R$ & $\at$ & 
$V$ (\%)& 
Pairs \\ 
& & & $\pm$ 0.1 & $\pm$ 0.1 & \\ \hline \hline
A & $<$ 21.775         & 1861 & -6.2  & 5.5  & 1,153,484 
\\ \hline
B & 21.775-47.925  & 1477 & -4.2  & 4.4  & 1,153,382 
\\ \hline
C & 47.925-82.550 & 1439 & -4.0  & 4.2  & 1,152,928 
\\ \hline
D & 82.550-130.200    & 1485 & -3.7  & 4.1 & 1,152,548
\\ \hline
E & 130.200-196.200    & 1461 & -4.0  & 4.0 & 1,152,448 
\\ \hline
F & 196.200-290.350    & 1406 & -3.8  & 4.0 & 1,152,440 
\\ \hline
G & 290.350-433.625& 1490 & -3.9  & 4.2 & 1,152,501 \\ \hline
H & 433.625-691.000& 1501 & -4.4  & 4.5 & 1,152,427 \\ \hline
I & $>$ 691.000        & 1883 & -5.3  & 5.2 & 1,149,599 \\ \hline
\end{tabular}
\caption{\label{tab:separation} Track separation squared for
each pair of multiple-muon tracks, divided into nine regions of
equal statistics, A...I.
The uncertainties on $\at$ and $V$ are from the fit.
}
\end{table}

\par
\begin{figure}[thbp]
  \begin{center}
    \includegraphics[width=0.48\textwidth]{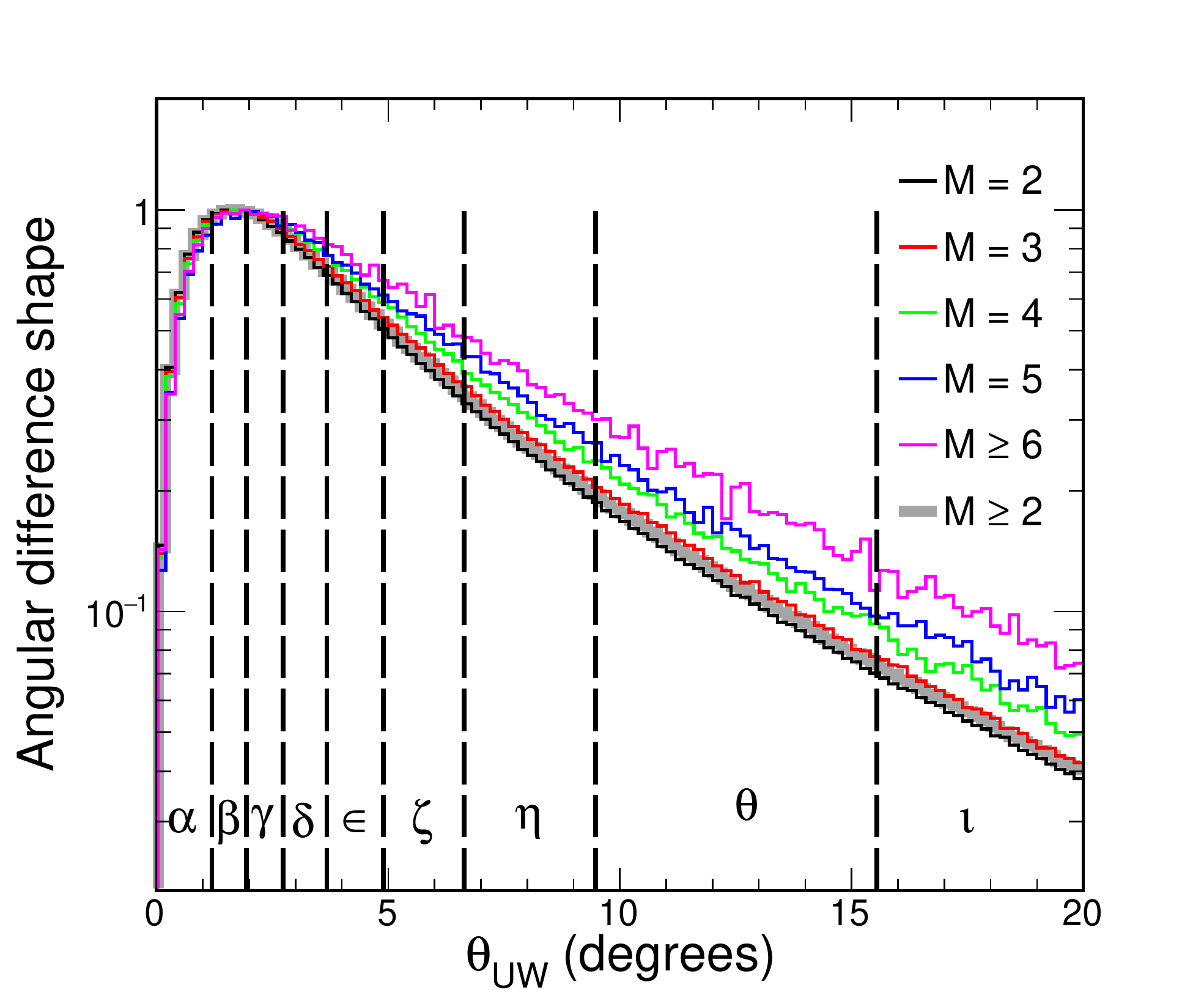}
  \end{center}
  \caption{Angle between each track in each muon pair in a multiple-muon 
event
in the NOvA ND.
Multiplicity distributions are normalized to have the maximum equal to 
1 (one).  The regions marked $\alpha$ to $\iota$ have equal statistics.
  \label{fig:angle}}
\end{figure}

 The angle between tracks in a multiple-muon event is also
related to the original muon energies.  For this we compute
\begin{equation}
\theta_{UW} = \arccos 
\left( 
\frac {{\vec{U}} {\cdot} {\vec{W}}} {|\vec{U}||\vec{W}|}
\right),
\end{equation}
where $\vec{U}$ and $\vec{W}$ are vectors representing the
directions of each pair of tracks in every multiplicity event.
Track angles 
may diverge
due to $p_t$ in the first interaction, different locations for
vertices in
further interactions, multiple scattering, and magnetic bending.
All of these effects are expected to be smaller for muons from higher 
energy 
primary
cosmic rays.  The angular resolution,
which is a function of
track length in the detector, affects this measurement.
From a MC simulation of 
parallel
tracks in the detector, the angular resolution for tracks which
enter the top and exit the bottom is 1.6$^{\circ}$.  The
distribution for the angle between all track pairs is shown in
Fig.~\ref{fig:angle}.  
\par The track angle data were divided into nine equal samples 
($\alpha$...$\iota$).  The seasonal 
parameters for these nine regions of angular separation are shown
in Table~\ref{tab:angle}.  There is a possible reduction in the
seasonal effect in the largest angle ($\iota$) bin.
We estimate a background of 600 two-muon events in two years from a
coincidence of two random single-muon events within 100$\,$ns, most
of which will be in the $\iota$ region $\theta_{UW}$ $>$ 15.55$^{\circ}$.
This background causes a negligible systematic 
uncertainty to 
our fits.  Another background which might contribute to the $\iota$ bin
is hadronic interactions just above the detector.

\begin{table}[thb]
\begin{tabular}{crrccr}
\hline
Sample & $\theta_{UW}$ (degrees) & $P_R$ & $\at$
& $V$ $\pm$ 0.1 (\%)& 
Pairs \\ 
\hline \hline
$\alpha$ & $<$ 1.19    & 644 & -4.2 $ \pm$ 0.1 & 4.6 & 1,206,534 \\ \hline
$\beta$ & 1.19-1.95    & 566 & -4.0 $ \pm$ 0.1 & 4.3 & 1,206,007 \\ 
\hline
$\gamma$ & 1.95-2.74   & 607 & -4.2 $\pm$ 0.1 & 4.4 & 1,207,553 \\ \hline
$\delta$ & 2.74-3.68   & 571 & -4.1 $\pm$ 0.1 & 4.4 & 1,206,531 \\ \hline
$\varepsilon$ & 3.68-4.90 & 582 & -4.3 $\pm$ 0.1 & 4.4 & 1,206,217 \\ 
\hline
$\zeta$ & 4.90-6.64    & 541 & -4.0 $\pm$ 0.1 & 4.3 & 1,206,216 \\ \hline
$\eta$ & 6.64-9.48     & 590 & -4.2 $\pm$ 0.1 & 4.5 & 1,206,275 \\ \hline
$\theta$ & 9.48-15.55  & 593 & -4.4 $\pm$ 0.1 & 4.6 & 1,206,196 \\ \hline
$\iota$ & $>$ 15.55    & 332 & -3.6 $\pm$ 0.2 & 3.5 & 1,200,193 \\ \hline
\end{tabular}
\caption{\label{tab:angle} Measurements of the seasonal
variation for
nine regions of angular separation of each muon pair in
a multiple-muon event
in the NOvA ND.
}
\end{table}
\par The muon multiplicity for multiple-muon events is a strong
function of the primary cosmic ray energy.  However, whatever
dynamics are controlling the seasonality of multiple-muon
events could be compounded as the multiplicity increases.
\par
\begin{figure}[thbp]
  \begin{center}
    \includegraphics[width=0.48\textwidth]{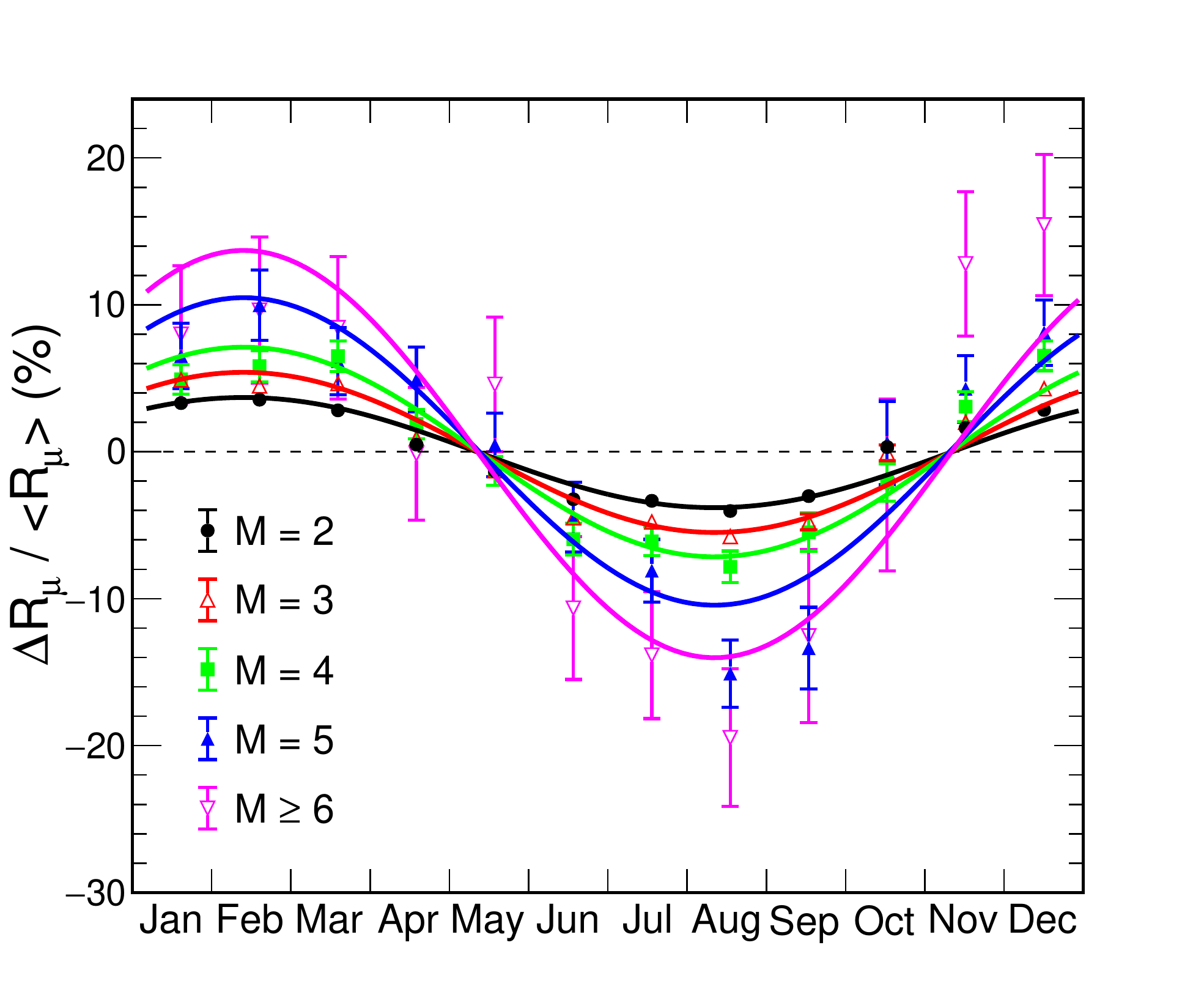}
  \end{center}
  \caption{Multiple-muon rate variation in the NOvA ND as a function of
month of year, shown for each multiplicity.  A cosine fit for each
data sample is also shown, representing an increase in the size of 
the
seasonal effect for larger multiplicities.  In each fit, the phase $\phi$ 
is fixed
to the value from the global fit.
  \label{fig:ampmult}}
\end{figure}

\begin{figure}[thbp]
  \begin{center}
    \includegraphics[width=0.48\textwidth]{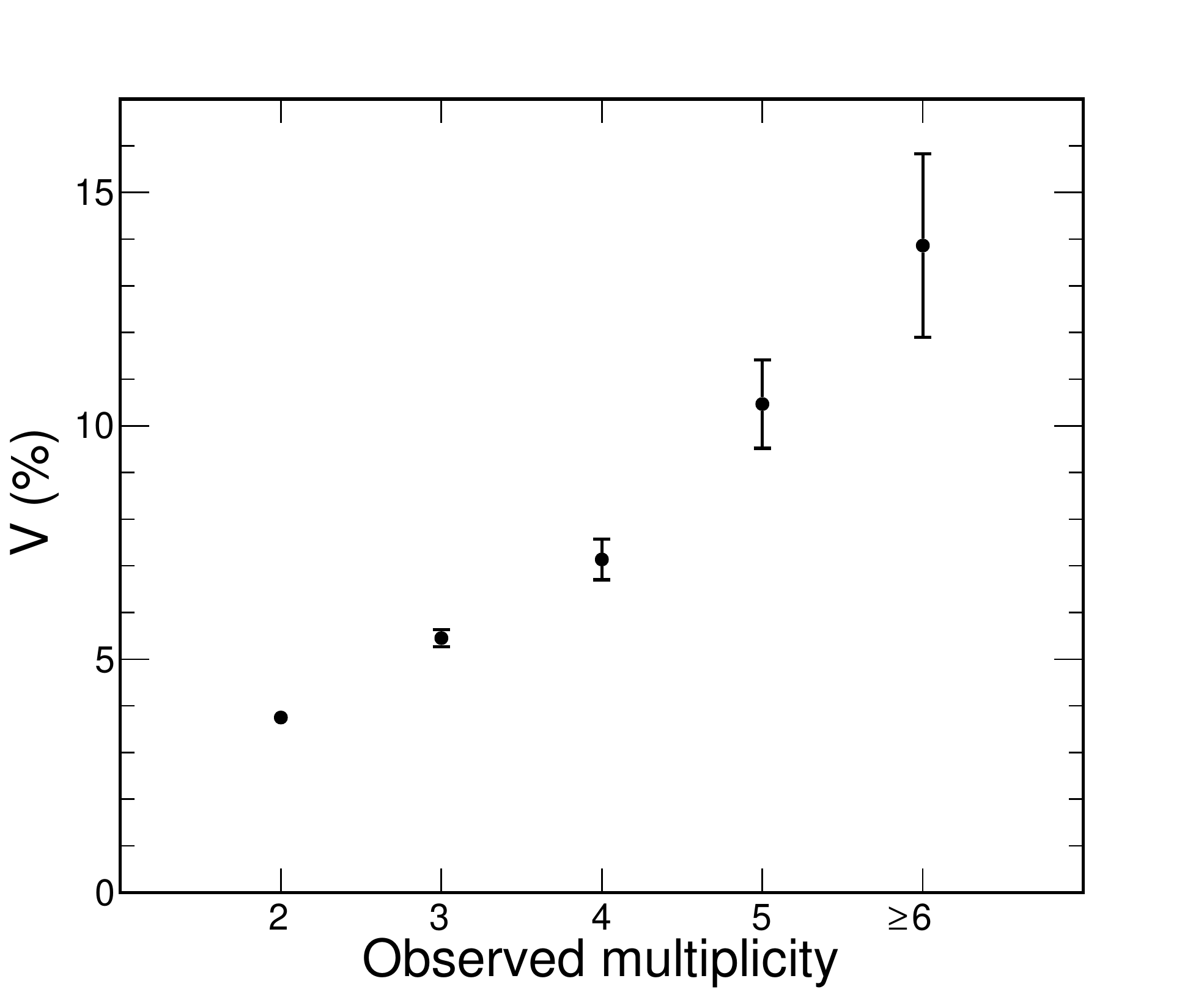}
  \end{center}
  \caption{Fitted amplitudes (\%) to a cosine fit of the seasonal
variation for each observed  
multiplicity
in the NOvA ND.  
  \label{fig:multfit}}
\end{figure}

 Since the statistics for each multiplicity are quite 
different, the Rayleigh power is not calculated.  Also, $\at$ is
not used since $\teff$ is multiplicity dependent in an unknown way.  
The amplitude fit for each multiplicity is shown in Table~\ref{tab:amult}.
The results of fitting the data to a cosine function
 for each multiplicity 
are shown in Fig.~\ref{fig:ampmult}.
A clear trend toward larger effects is seen as the multiplicity
grows.  The amplitude is shown as a function of multiplicity in
Fig.~\ref{fig:multfit}.

 \begin{table}[thb] \begin{tabular}{lccl} 
\hline Multiplicity   & $V$(\%)& & \\ \hline \hline
2 & 3.81 &$\pm$& 0.05 \\ \hline
3 & 5.5 &$\pm$& 0.2  \\ \hline
4 & 7.1 &$\pm$& 0.4  \\ \hline
5 & 10.0 &$\pm$& 0.9 \\ \hline
$\ge$ 6 &14 &$\pm$& 2  \\ \hline \hline
$\ge$ 2 &4.1 &$\pm$& 0.2  \\ \hline
\end{tabular}
\caption{\label{tab:amult} Amplitude as a function of multiplicity.}
\end{table}

\section{Systematic Uncertainties}
Our conclusions involve the presence of a seasonal effect with
a maximum in the winter which grows with multiplicity, and the absence of 
a noticeable trend in the size of that effect for three other variables.
While there is no parameter for which a systematic uncertainty is 
appropriate,
we must be confident that no systematic effect could create or mask
the observed results.  The Rayleigh power gives a measure of the
statistical power of a periodic signal.  For every sample studied,
the Rayleigh power suggests the presence of a seasonal effect with a
truly negligible chance probability.
\par The fits to $\at$, which depends on temperature data,
and the amplitudes $V$ of the cosine fit, which do not depend on any
temperature data, give qualitative measurements of the size
of the seasonal variation which agree.  The individual
temperatures from ECMWF used to calculate $\teff$ have a systematic
uncertainty of $\pm0.31$ K \cite{bib:minosf}.  Based on the 
variation in
temperature over the longitudinal area contributing to observed
muon production, a systematic uncertainty on $\teff$ of 
$\pm$0.1 K
was determined \cite{bib:stefthesis}.  MINOS measured $\at$ for
single muons at this location to be +0.428 $\pm$ 0.003(stat)
$\pm$ 0.059(syst) \cite{bib:minosn}.  The positive 
value indicates a summer maximum.     
We calculated $\at$ for single muons in NOvA and
similarly found a summer maximum using
daily and weekly time bins, and due to the lack of
consistency in the value of $\at$ we assign a systematic
uncertainty of $\pm$ 0.4, which corresponds
to an uncertainty on the amplitude of $\pm 0.8\%$.
This number provides a maximum correlated 
systematic uncertainty for $\at$ for multiple muons
and includes all potential 
effects from
temperature measurements and hardware or reconstruction issues
which might be seasonally time dependent.  
Every measurement of $\at$ in 
Tables~\ref{tab:zenith}, \ref{tab:separation}, and \ref{tab:angle}
was negative with an absolute value at least 8 times this
systematic uncertainty, indicating an unambiguous winter maximum.
In the 
calculation of $\teff$, the weighting of the atmospheric
temperature versus altitude was done for the calculated location of
single-muon and not multiple-muon production.  The values of
$\at$ in this analysis should be interpreted as a parameter indicating
the size and sign of the seasonal effect, and not strictly the 
correlation coefficient between rate and an appropriately 
calculated $\teff$.
\par The deadtime of the activity trigger
  used to acquire the cosmic ray data is slightly higher
  in the winter than the summer at the sub-percent level, due to the
  NuMI beam schedule.  This deadtime difference based on
our monitoring could affect 
the value
of  $V$ by at most 0.5\% and has not been corrected.  This effect would
be included in the $\pm 0.8\%$ uncertainty from the single-muon $\at$
inconsistency and
could be the major contributor to it.   

\par While the average temperature per month does not strictly
follow a cosine curve, and hence its effect on seasonal variations
would not either, the data in Figs. 5 and 9 follow a cosine function
well 
enough
for a fit to the amplitude of a cosine function to give a reasonable
measure of the size of the seasonal variation.  
In order to evaluate the effect of the assumed shape of the distribution
on the amplitude of the fit, a new fit was made by choosing a
 correlated systematic uncertainty on the rate such that $\chi^2/$dof =1.
That new fit to the data in Fig.~\ref{fig:monthofyear} gave $V$ = 3.9 
$\pm$ 
0.4.
We interpret 0.4 as a potential deviation in the value of $V$ for
the fact that true seasonal variations in our data do not follow
a cosine.  All values of $V$ in
Tables~\ref{tab:zenith}, \ref{tab:separation}, and \ref{tab:angle}
are at least 8 times this deviation.

The reconstruction program that we used did not reconstruct all
triggered muons, particularly short and steep tracks.  The
inefficiency was not negligible.  Visual inspection and MC studies showed 
that
all reconstructed events were pure in the sense that there were
at least the identified number of throughgoing muons in each event.
For example, a reconstructed 3-muon event could possibly have 4 or
more throughgoing tracks, but not 0, 1 or 2.  
This reconstruction issue could decrease the apparent size of that
dependence but could not create a spurious dependence.  The known steep 
falloff in the true multiplicity 
distribution \cite{bib:kasahara} implies this uncorrected multiplicity 
distribution does not 
change our conclusion
that the seasonal effect grows with multiplicity.
The conclusion in the
paper, that there is a multiplicity dependence as indicated in 
Fig.~\ref{fig:multfit},
is robust.   

\par We have not identified any systematic uncertainty which depends
strongly on spatial separation, angular separation, or
zenith angle.
The systematic uncertainties involving deadtime and temperature
cancel
to first order when dividing the data into bins of
these observables and
do not mask the lack of trends in 
Tables~\ref{tab:zenith}, \ref{tab:separation}, and \ref{tab:angle}.

\section{Summary and discussion}

The NOvA ND data show that the rate of multiple muons seen at a depth of 
225$\,$mwe underground is anticorrelated with the temperature of the 
atmosphere. That is, the rate increases in the winter and decreases in the 
summer. This anticorrelation between temperature and rate was also 
observed previously \cite{bib:minos}. 
\par In this analysis we used several 
proxies for the initial muon and primary cosmic ray energies to see if the 
effect was related to the particle initial energy; there is no indication 
that is the case. However, we observe the effect grows from 4\% 
to 14\% 
with increasing muon multiplicity. 
This is a new observation, which may 
allow one to clarify further the physics origin of the observed puzzling 
behavior.
The quantitative nature of this 
anticorrelation is not understood. This result is consistent with the 
suggestion from the previous analysis in which the effect is attributed to 
multiple muons coming from those $\pi$'s which are more likely to interact 
than decay in the winter \cite{bib:minos}. Thus the single-muon rate is 
higher in the 
summer and the multiple-muon rate is higher in the winter. 
\par The mean surface muon energy for muons reaching a depth of 225$\,$mwe 
is 
below the critical energy for both $\pi$'s and K's, so that more 
secondary 
hadrons are decaying before they interact in the upper atmosphere. For 
detectors at depths of 2000 mwe or more, the mean muon energy is above the 
critical energy for $\pi$'s and comparable to the critical energy for 
K's. 
The observed effect at 2000 mwe is more complicated than just a dependence 
on the $\pi$ and K critical energies and so further studies should be 
done at those depths. The results from the NOvA ND  presented in this 
paper will be important inputs to future simulation and study of this 
effect.

\section{Acknowledgments}
\label{sec:Acknowledgements} 

This document was prepared by the NOvA collaboration using the resources 
of the 
Fermi National Accelerator Laboratory (Fermilab), a U.S. Department of 
Energy, 
Office of Science, HEP User Facility. Fermilab is managed by Fermi 
Research 
Alliance, LLC (FRA), acting under Contract No. DE-AC02-07CH11359.
This work was supported by the U.S. Department of Energy; the
U.S. National Science Foundation; the Department of Science and
Technology, India; the European Research Council; the MSMT CR, GA UK,
Czech Republic; the RAS, RFBR, RMES, RSF, and BASIS Foundation,
Russia; CNPq and FAPEG, Brazil; STFC, and the Royal Society, United
Kingdom; and the state and University of Minnesota.  We are grateful for 
the 
contributions of the staffs of the University of Minnesota at the Ash 
River Laboratory and
of Fermilab.

\end{document}